\documentclass[conference]{IEEEtran}
\IEEEoverridecommandlockouts % footnote
\usepackage{blindtext, graphicx}

\usepackage{verbatim} %% comment command
\usepackage{amsmath}
\usepackage{amsfonts}
\usepackage{subfigure}
\usepackage{float}  % 
\usepackage{color}

\newcommand{\n}{|\!|}
\ifCLASSINFOpdf
\else
\fi
\graphicspath{{figures/}{figures2/}}

\begin{document}
\title{Quantifying Volatility Reduction \\in German Day-ahead Spot Market \\
in the Period  2006 through 2016
\thanks{This work was supported in part by the Dutch STW under project grant \emph{Smart Energy Management and Services in Buildings and Grids (SES-BE)}.}
}

\author{\IEEEauthorblockN{Abdolrahman Khoshrou}
\IEEEauthorblockA{Intelligent and Autonomous Systems Group\\
Centrum Wiskunde \& Informatica\\
The Netherlands\\
Email: a.khoshrou@cwi.nl}
\and
\IEEEauthorblockN{Eric J. Pauwels}
\IEEEauthorblockA{Intelligent and Autonomous Systems Group\\
Centrum Wiskunde \& Informatica\\
The Netherlands\\
Email: Eric.Pauwels@cwi.nl}
}

%Telephone: (800) 555--1212, Fax: (888) 555--1212}


\maketitle
\pagestyle{empty} % plain empty
\begin{abstract}
%\boldmath
In Europe, Germany is taking the lead in the switch from the conventional to renewable energy. 
This poses new challenges as wind and solar energy are fundamentally intermittent, weather-dependent and less predictable. 
It is therefore of considerable interest to investigate the evolution of price volatility in this post-transition era.
There are a number of reasons, however, that makes the practical studies difficult. 
For instance, EPEX prices can be zero or negative. 
Consequently, the standard approach in financial time series analysis to switch to logarithmic measures is inapplicable.
Furthermore, in contrast to the stock market prices which are only available for trading days, EPEX prices cover the whole year, including weekends and holidays. 
Accordingly, there is a lot of underlying variability in the data which has nothing to do with volatility, but simply reflects diurnal 
activity patterns.
An important distinction of the present work is the application of matrix decomposition techniques, namely the singular value decomposition (SVD), for defining an alternative notion of volatility.
This approach is systematically more robust toward outliers and also the diurnal patterns.
Our observations show that the day-ahead market is becoming less volatile in recent years.
\end{abstract}
%what effect these changes have on the volatility of the electricity price. 
%%%While market coupling serves as a way to reduce price volatility, dependence on intermittent renewable energy sources (RES) might have the opposite effect. 
%To elucidate the combined impact of these two developments, we investigate the evolution of the electricity price on the German EPEX day-ahead spot market from 2006 to 2016. 
\begin{IEEEkeywords}
Day-ahead price,
Electricity market,
Energy switch,
Singular value decomposition,
Time-series analysis,
Renewable energy sources.
\end{IEEEkeywords}
% \fi
\IEEEpeerreviewmaketitle
%%%%%%%%%%%%%%%%%%%%%%%%%%%%%%%%%%%%%%%%%%%%%%%%%%%%%%%%%
\section{Introduction}
Renewable Energy Sources (RES) are assuming an increasingly pre-eminent role in the German electricity production.
Significant cost reductions, on the one hand, and tremendous technology advances and reliability improvements, on the other hand, have primed a growing interest in  green electricity.
Germany is pursuing an ambitious goal, a switch from fossil fuel to renewables, ``Energiewende'' (energy transition). 
By 2050, the emission of greenhouse gases is planned to reduce by 80-95\% (see~\cite{energiewendeReport}).
To achieve this, energy consumption is to be reduced by 50\% and at least 80\% of electricity is to come from renewables.
In line with that, Germany has substantially expanded its RES-capacity, in particular wind and solar~\cite{GE_wind}.
Consequently, the need for accurate predictions for the quantity of green electricity which is going to be fed into the grid at any given moment is becoming increasingly important.
Moreover, energy transition could put pressure on the energy sector in terms of flexibility in managing power demand and supply. 
It is therefore to be expected that this evolution will have a significant impact on German electricity prices.
%This impact could be complex because two opposite forces are at play.
%On the one hand, the marginal cost of RES is relatively low (especially if subsidized) and increased penetration of wind and solar would therefore induce a downward trend in electricity prices. 
%On the other hand, wind and solar are intermittent and fluctuating energy sources and the associated uncertainty is expected to push the prices higher. Indeed, if an abundance of German renewables substantially impacts the price and volatility of the electricity in Germany, it affects the position of fossil fuel suppliers.

This paper provides an overview of the price volatility of the German day-ahead EPEX market.   
If price volatility increases, it can cause additional risks for suppliers and consumers on the electricity market. 
Negatively impact the reliability of the power grid is also possible. 
Due to the integration of the European grids the problems will not be limited to the German grid. 
Increasing instability in the German grid means also a higher instability in the neighboring grids.
In the present work, we have opted to focus on this market as it represents an important and growing segment where market mechanisms are clearly visible.  
In particular, we focus on the following question: 
How can the evolution of the price volatility of the day-ahead market over the past eleven years (i.e., 2006-2016) be quantified?
%If the accuracy of the renewable power generation forecasting tools have increased substantially over the recent years?
%Do we see their effect on the price volatility of the  day-ahead  market over the past eleven years (i.e. 2006-2016)? How can we quantify the volatility? 

The rest of this paper is organized as follows: Section~\ref{sec:background} contains background studies and literature review. Section~\ref{sec:data} focuses on the data description for day-ahead market; also explains the source and a brief summary of the day-ahead market mechanism.
The evolution of a new type of volatility measure for time series data is discussed in Section~\ref{sec:methodology}.
The general reduction in the volatility of price data is argued in Section~\ref{sec:conclusion}.
%%%%%%%%%%%%%%%%%%%%%%%%%%%%%%%%%%%%%%%%%%%%%%%%%%%%%%%%%%%%%%%%%%%
\section{Background and Literature Review} \label{sec:background}
%%%%%%%%%%%%%%%%%%%%%%%%%%%%%%%%%%%%%%%%%%%%%%%%%%%%%%%%%%%%%%%%%%%
Recently, the impact of variable generation on the electricity market has attracted a lot of attention.  
We briefly highlight a number of important contributions which are related to the topics discussed in this paper. 
Denny et al.~\cite{denny2010impact} study the functionality of the increased interconnection between Great Britain and Ireland to facilitate the integration of the wind farms into the power system. 
This work suggests a reduction in average price and its volatility in Ireland to be an outcome of the increased interconnection.
With the growing contribution of intermittent energy sources, transmission grid extensions and increasing the cross-border interconnection capacities seem inevitable.
Schaber et al.~\cite{schaber2012transmission} examine the viability of this approach and its effects, based on projected wind and solar data until 2020; they also conclude that expanding the grid is, indeed, helpful in coping with externalities which come with the deployment of RES.  
The relation of substantial expansion of photovoltaic (PV) installations in Germany and Italy with daytime peak price fall in these countries is discussed in~\cite{barnham2013benefits}. 
In~\cite{adaduldah2014influence} the influence of the RES on the German day-ahead market is studied; the authors also have considered priority that the German government assigns to the green electricity over fossil fuels in case of adequate supply.  
This paper also argues that the emergence of negative prices in the German day-ahead market is the result of the integration of RES.

In the present paper, our main focus is the evolution of the price volatility.
Therefore, a novel approach to quantify the volatility of the German day-ahead market is proposed.
% What has attracted less attention, however, is the volatility of the market.
%%%%%%%%%%%%%%%%%%%%%%%%%%%%%%%%%%%%%%%%%%%%%%%%%%%%%%%%%%%%%%%%%%%
\section{Data} \label{sec:data} % European Power Exchange EPEX SPOT
\begin{figure}
  \begin{center}
    \leavevmode
    \includegraphics[width=8cm,height=5cm]{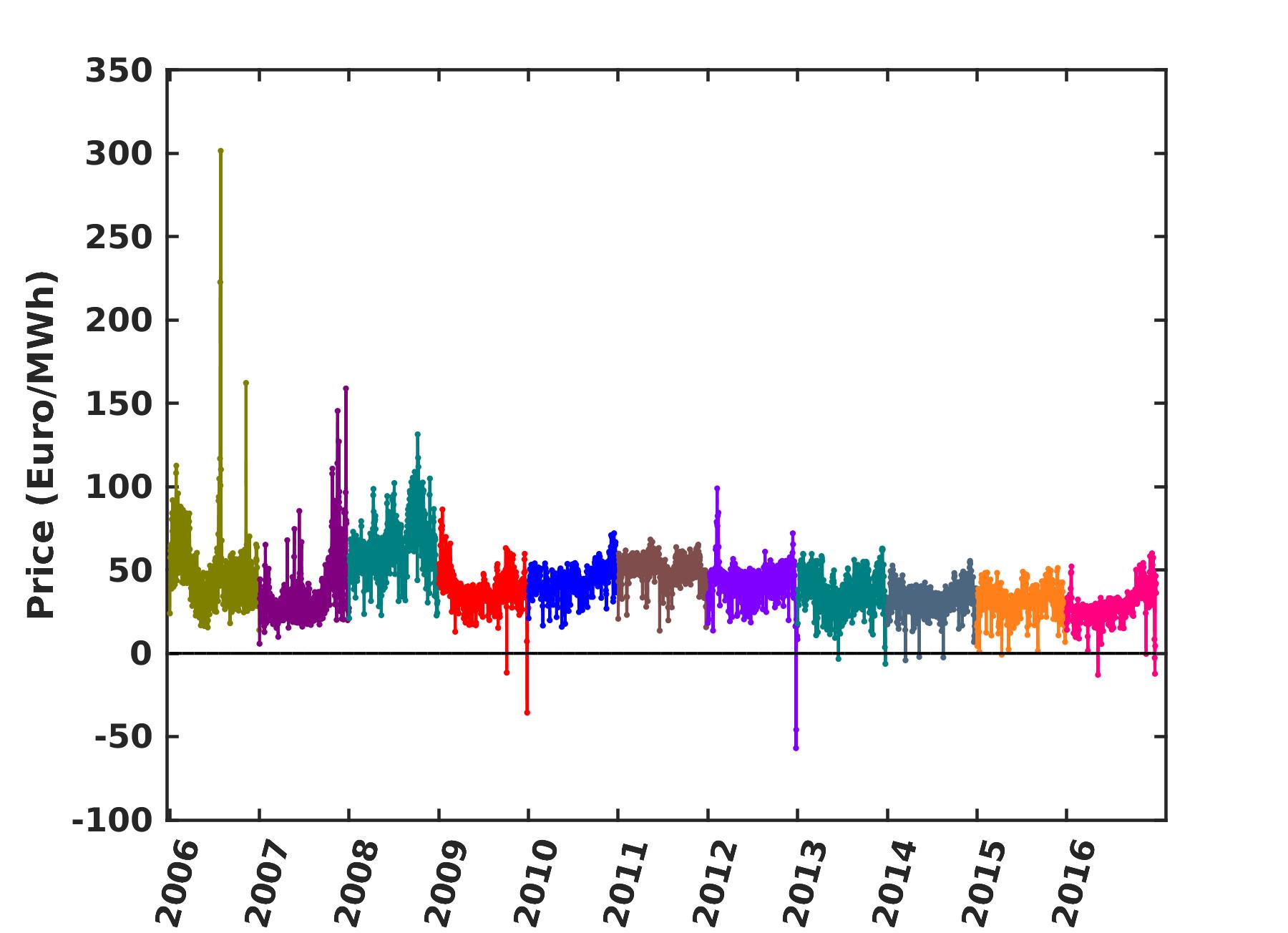}
    \caption{
    German day-ahead spot prices from 2006 through 2016 (daily averages); an overall downward trend in the day-ahead price is notable. }
    \label{fig:main_270916_GE_draft_fig8}
  \end{center}
\end{figure}

%%%%%%%%%%%%%%%%%%%%%%%%%%%%%%%%%%%%%%%%%%%%%%%%%%%%%%%%%%%%%%%%%%%
The European Power Exchange (EPEX SPOT SE) operates on the Central Western European (CWE) spot market, i.e. Swiss, French, German and Austrian short-term electricity markets.
Striving for the creation of a single integrated electricity market, EPEX SPOT functions as an organized wholesale market place for trading large quantities of electricity between the market members.
These members are mostly non-final consumers and big players in the energy sector such as utilities and aggregators, industrial producers, the Transmission System Operators (TSOs), banks, financial service providers and energy trading entities that are working within the energy sector on a daily basis.
In fact, this company offers its clients the technologies, electronic trading systems and platform to operate their orders based on reference prices.
% to save some space I put them as comments
\begin{comment}
Energy trading entities, banks and financial services providers play a prominent role in increasing liquidity of the wholesale power market. 
These members are mainly focused on market and trade cross borders, even though not necessarily owning any power assets.
Some energy intensive industries also participate in the wholesale market as bargain hunters to get a deal at the most reasonable price. 
Grid losses compensation is a great prime for TSOs to intervene on the spot market. 
Furthermore, regulating feed-in tariff schemes for marketing zero-carbon energy sources are extensively practiced by the TSO in Germany.
Since  EPEX SPOT markets allow for proxy trading, several members trade on behalf of smaller companies or aggregators that pool small assets. 
This is important since these small entities are mostly formed by decentralized green electricity producers (mainly hydro in France, wind and solar in Germany), demand response aggregators, or other very small suppliers, for whom direct trading would be too costly or cumbersome.  
\end{comment}
\paragraph{Day-ahead Auction Spot Market}
The day-ahead market is an exchange for short-term electricity contracts. 
The trading in this market is driven by its participants.
A buyer, typically a utility, needs to assess how much energy (MWh) it will need to fulfill its customers 
requirements for the following day, and how much its purchase price is going to be (Euro/MWh) hour by hour. 
The seller, for example the owner of a wind farm, also submits the quantity he is prepared to deliver the next day and the price level on hourly basis. 
The deadline for the members to submit the price and the quantity for which they seek to make an agreement 
is 12:00 CET.
These ``bids'' are fed into a complex algorithm to calculate the \textit{clearing} price. 
%This process typically takes around 42 minutes to clear the market and settle the financial and physical transactions.
From 00:00 CET the next day, the sellers deliver the power at the contracted rate.
%In this paper hourly day-ahead spot market data in~\cite{epexPrice} is used.
\paragraph{Day-Ahead Spot Prices (in Euro/MWh)}
This is the price (for each time-slot of the next day) as set by the spot market. 
As previously mentioned, on the day-ahead market the hourly price of the traded quantity (in Euro/MWh) is set a day earlier. Fig.~\ref{fig:main_270916_GE_draft_fig8} illustrates an overview of the  hourly values of the price on the day-ahead market in Germany and Austria from 2006 until 2016 (data source:~\cite{epexPrice}).
%%%%%%%%%%%%%%%%%%%%%%%%%%%%%%%%%%%%%%%%%%%%%%%%%%%%%%%%%%%%%%%%%%%%%%
\section{Volatility of the Day-Ahead Price} \label{sec:methodology}
%%%%%%%%%%%%%%%%%%%%%%%%%%%%%%%%%%%%%%%%%%%%%%%%%%%%%%%%%%%%%%%%%%%%%%
\subsection{Introduction}
It is reasonable to question whether the intermittency of renewables would render the price more erratic. 
This can happen due to technical problems in the grid, e.g., congestion; or simply due to poor performance of the day-ahead energy forecasting models.
For that, we are interested in the volatility of the day-ahead price. 
Loosely speaking, volatility refers to the random fluctuations of a time series about its  expected value. 
There are various methods to define and quantify volatility, from applied models like Garman/Klass to coefficient of variation and formal Stochastic Volatility models such as GARCH, Heston models and 
the like, see, e.g.,~\cite{rogers1991estimating,baillie1996analysing}.
However, there are at least two reasons why it is problematic to blindly transfer standard fintech methodology to the current setting: 
\begin{itemize}
\item[$\bullet$] Since EPEX prices can be zero or negative, the standard approach to switch to logarithmic measures is not applicable.
\item[$\bullet$] More importantly, while the stock market prices are only available on trading days; EPEX prices cover the whole year, including weekends and holidays. As a consequence, there is a lot of underlying variability in the data which has nothing to do with volatility, but simply reflects the diurnal activity patterns. 
\end{itemize}
For these reasons, we will first pre-proccess the raw data to eliminate the underlying patterns; and subsequently focus on quantifying the volatility of the resulting {\it residuals}.  The nature of this pre-processing is discussed in the following subsection. 
\subsection{Extracting underlying trends}
\begin{figure}
\centering
  \includegraphics[width=4.15cm,height=3cm]{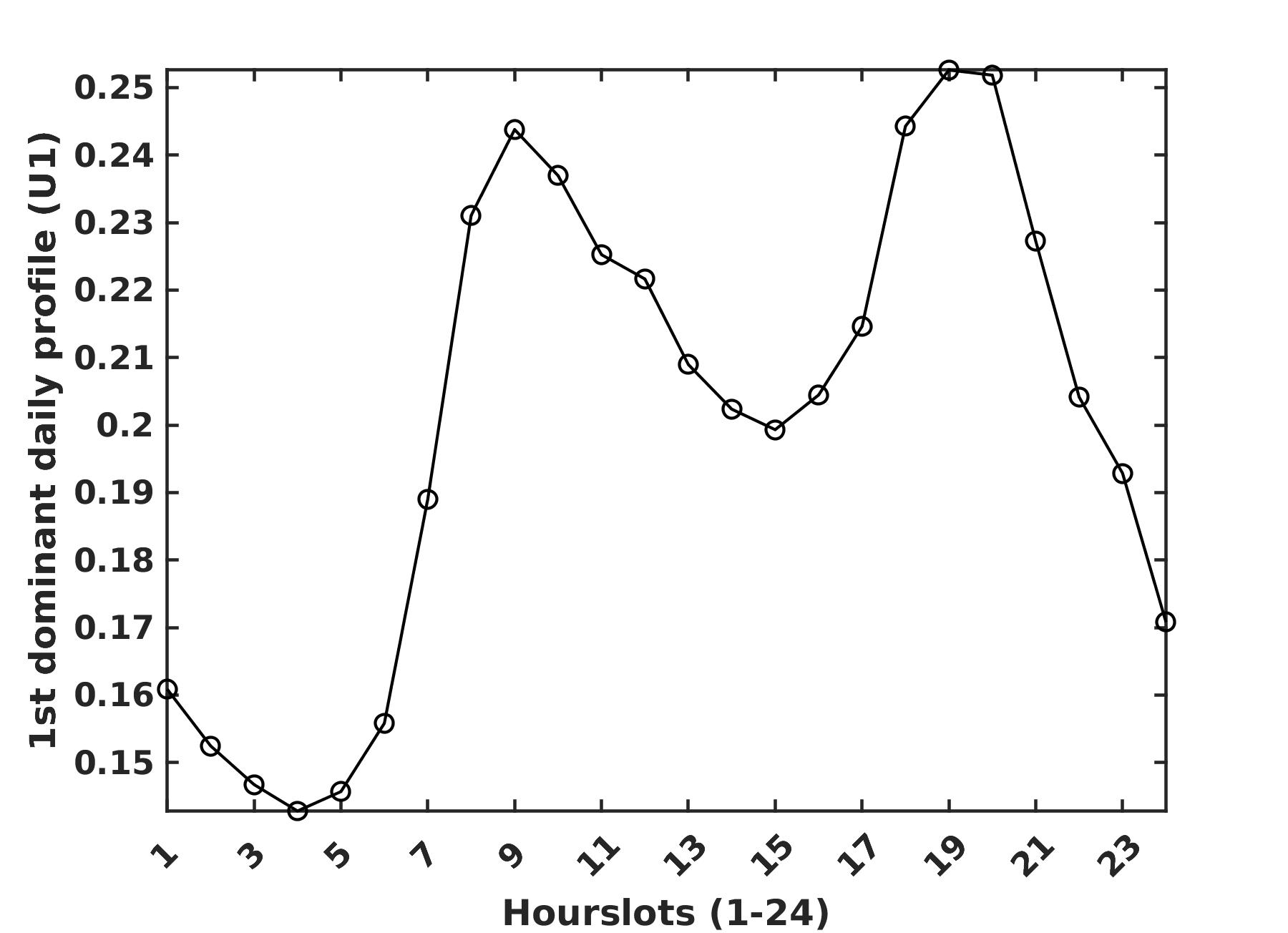}
  \includegraphics[width=4.15cm,height=3cm]{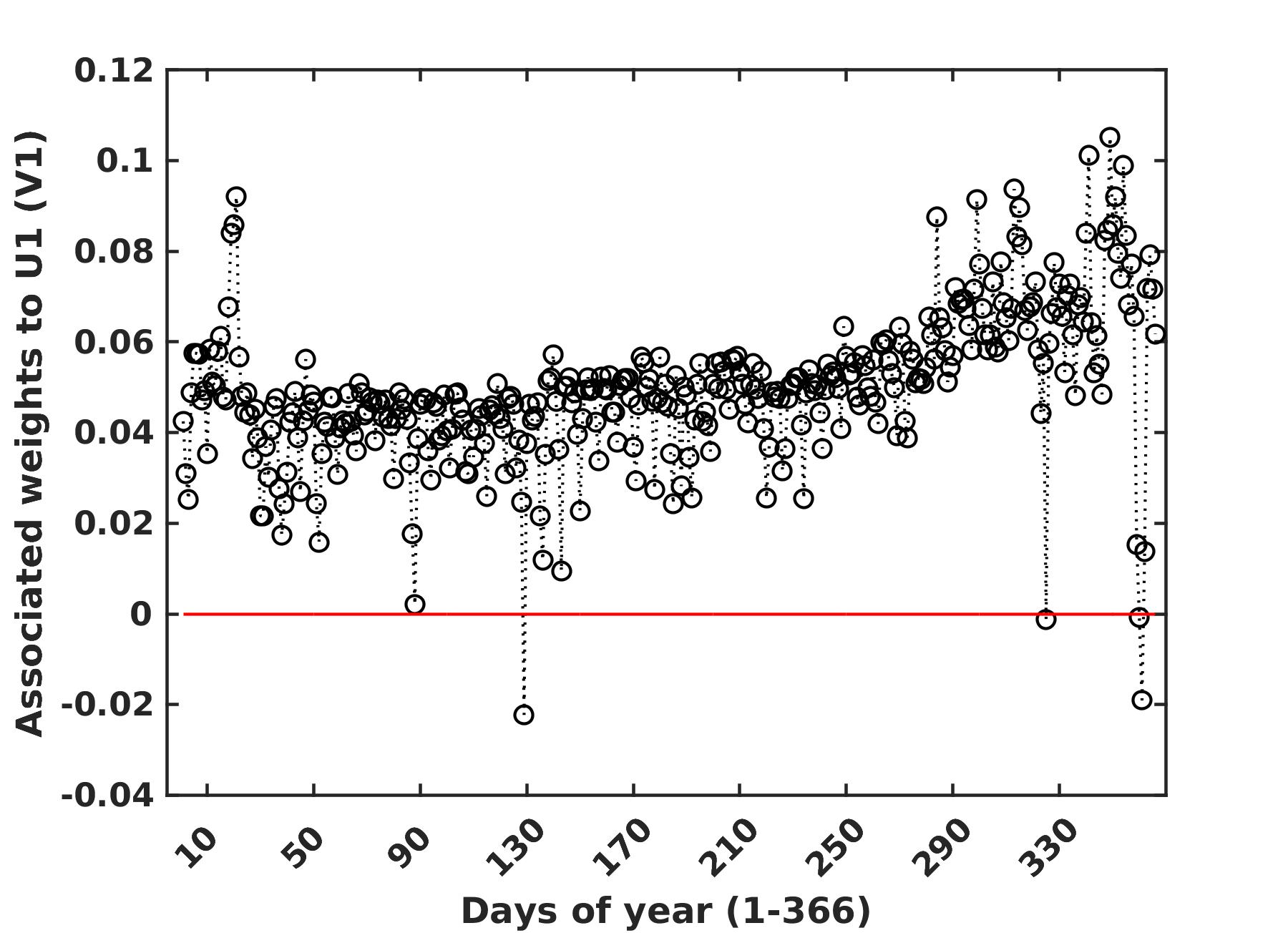}\\
  \vspace{.25cm}
  \includegraphics[width=4.15cm,height=3cm]{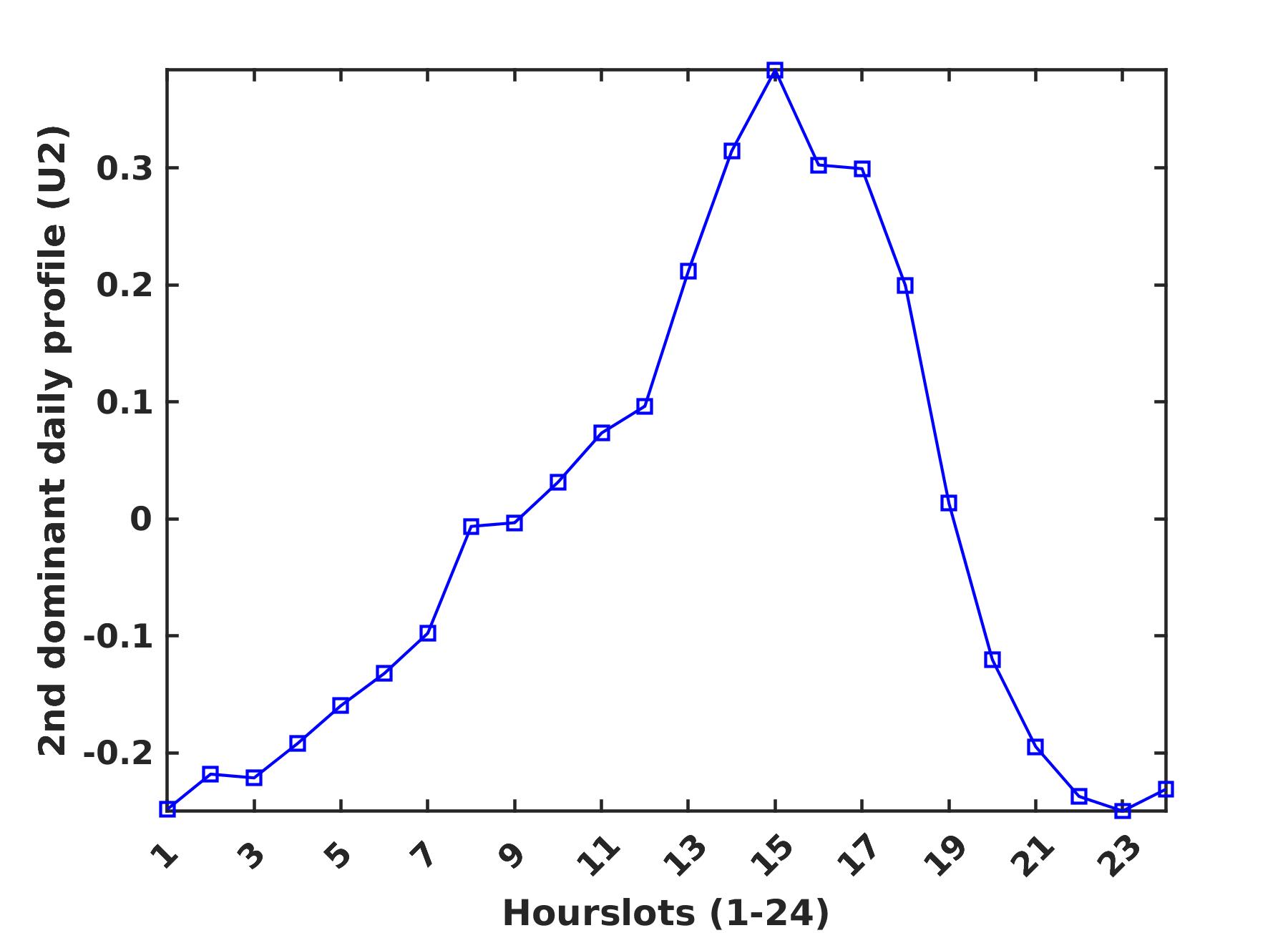}
  \includegraphics[width=4.15cm,height=3cm]{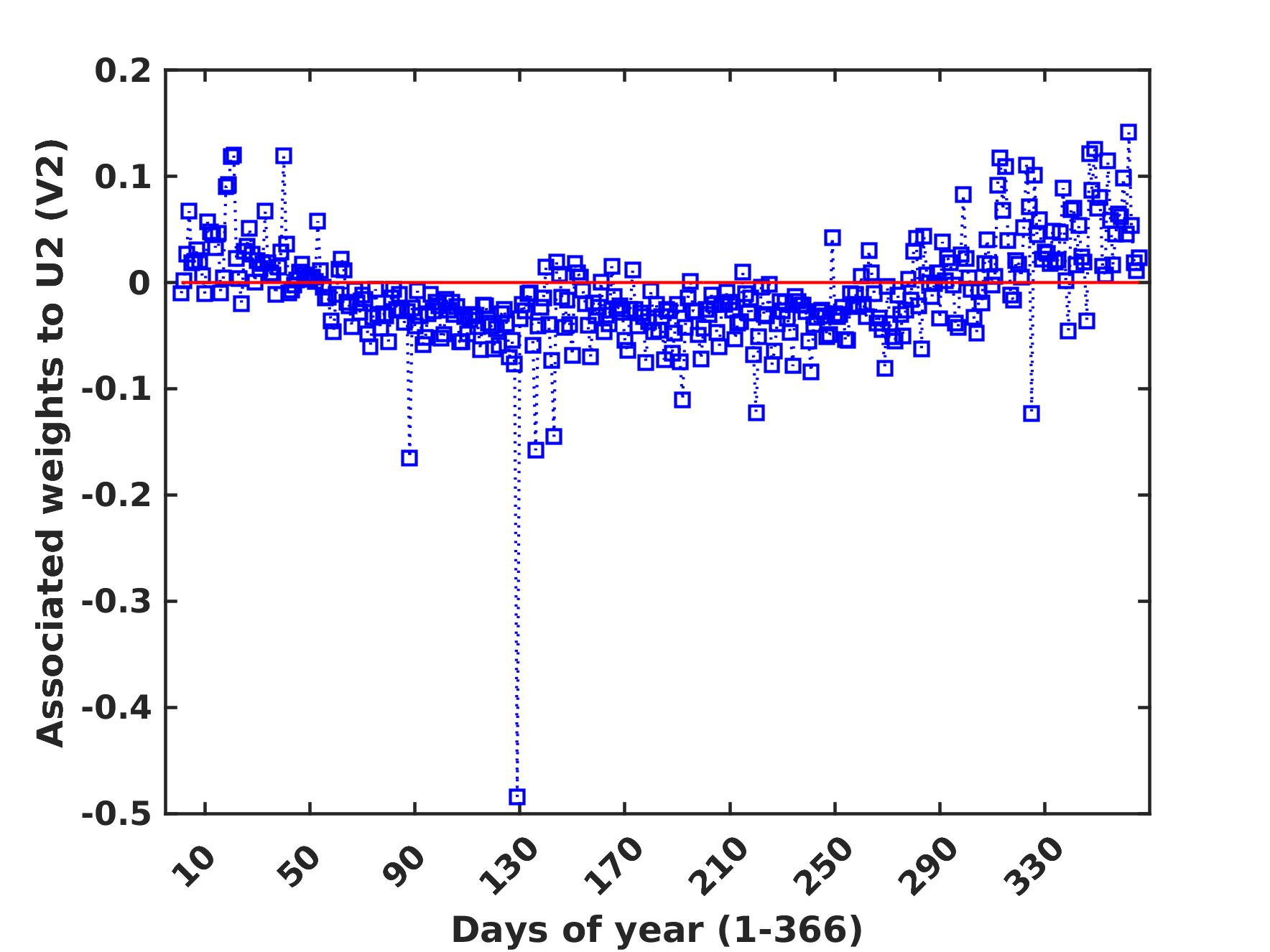}
  \caption{SVD-based rank-2 approximation:   Left column: first two columns of U-matrix representing a weighted averaged profile for each day (top) and a first order correction (again one value for each hour over a 24 hour period). Right column: corresponding amplitudes (V columns).}
  \label{fig:svd_rk2_approx}
\end{figure}

\begin{figure}
  \centering
  \includegraphics[width=7.1cm,height=3.5cm]{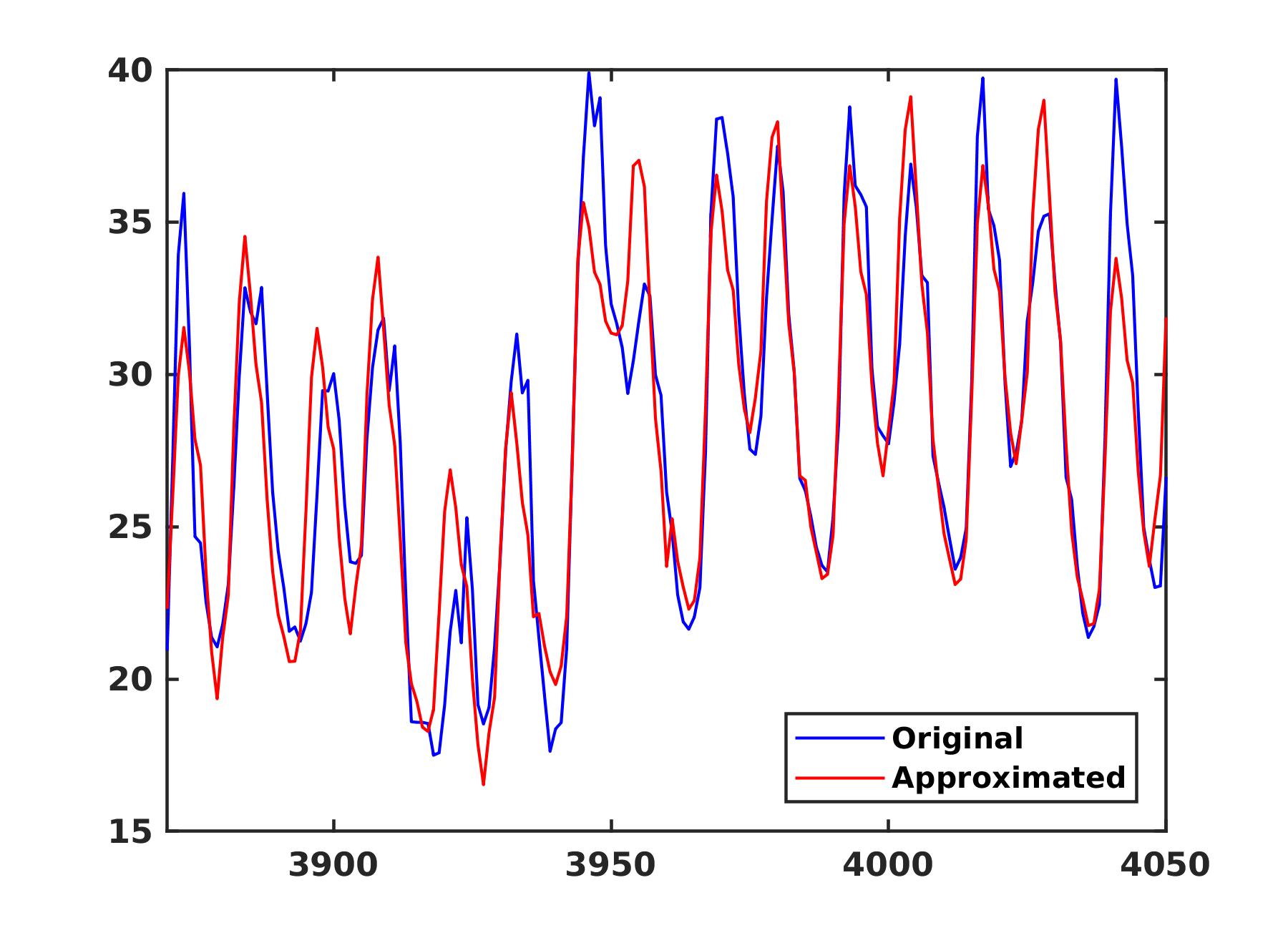}
  \caption{Detail of rank-2 approximation (red) superimposed on original data (blue).}
  \label{fig:svd_rk2_approx_detail}
\end{figure}

\begin{figure}[]
\centering
\hspace{-.4cm}
 \includegraphics[width=7.2cm,height=3cm]{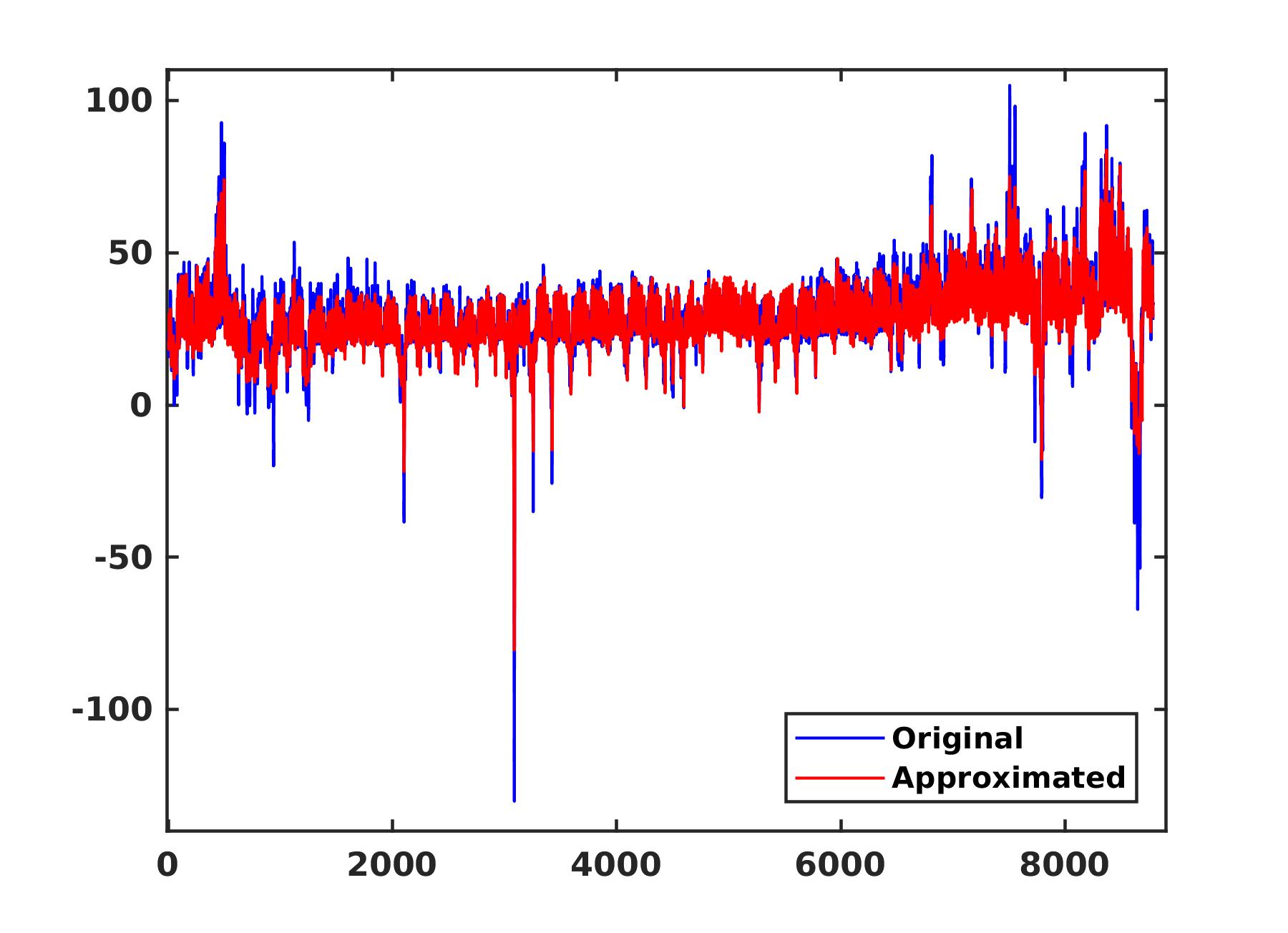}
  \includegraphics[width=7cm,height=3cm]{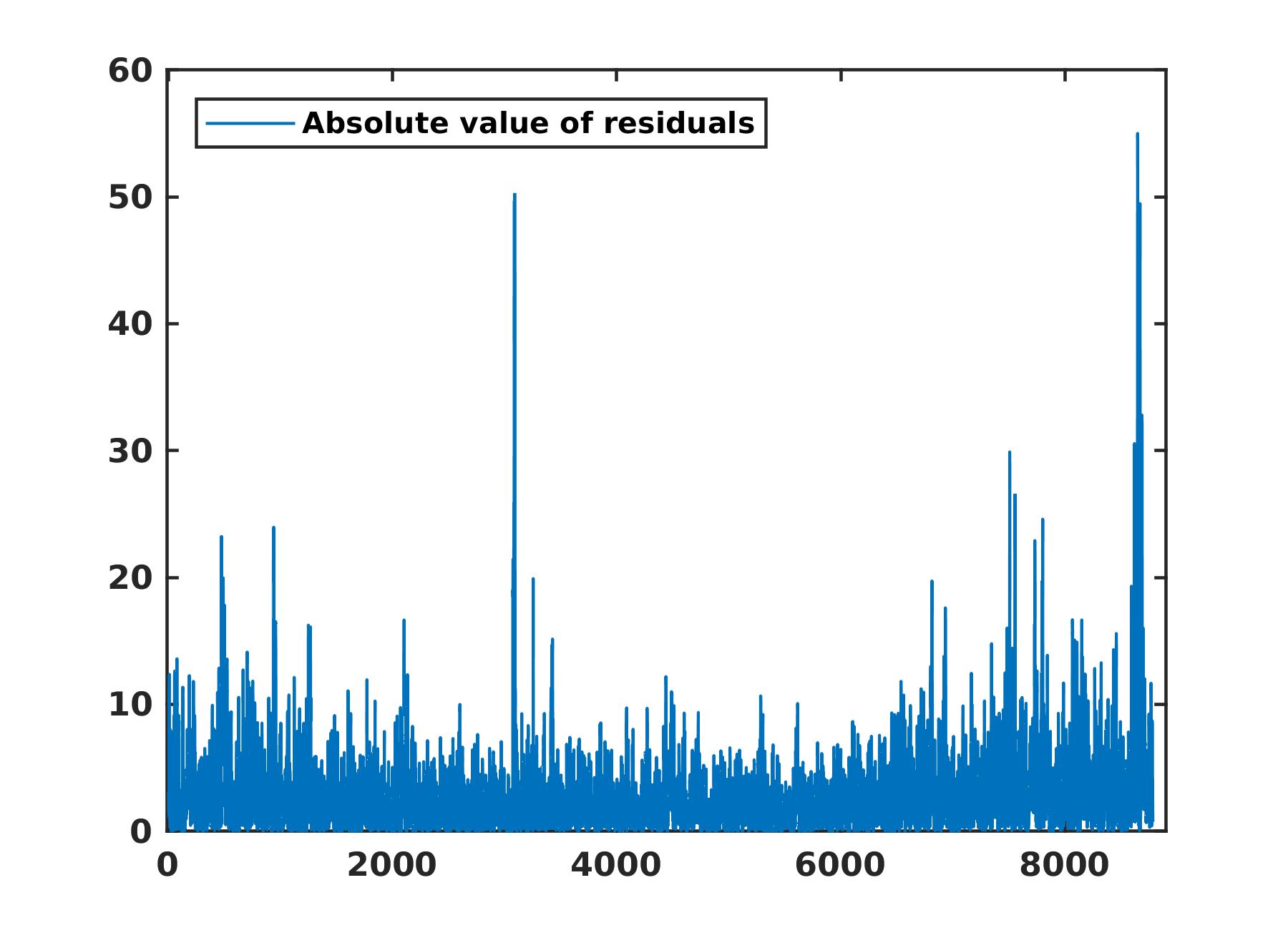}
  \caption{Top: Original and rank-2 approximation of the price data for 2016. Bottom: Absolute values of the residuals after rank-2 approximation. Residuals as a means to measure volatility.}
  \label{fig:2016_price_residual}
\end{figure}

Singular value decomposition (SVD) is a popular matrix decomposition technique; which we use to extract the underlying daily and seasonal patterns.
SVD is applied  extensively in matrix computations, but can also be put to good use in the study of time series which have exogenously induced periods. 
This is often the case in economics time series, where the variables of interest show daily or annual periods. 
In the case at hand, it is clear that prices would show daily patterns that might change relatively slowly over the year.
To apply SVD, we rewrite the time series as a matrix where each column records the 24 hourly values for a particular day~\cite{8260303}. 
In this way, a time series representing a typical year is recast as a matrix $A$ of size $24 \times 365$. 
If the results for every day were identical, then all the columns would be identical and the matrix would 
have a rank equal to one ($rank(A) =1$). Put differently, the matrix could be written as the product of 
a $24\times 1$ column and a $1\times 365$ row. 
In practice, however, the values for subsequent days will tend to be slightly different and therefore $A$ has a rank that exceeds one. 
Nevertheless, SVD assures us that $A$ can still be expanded as a sum of rank 1 matrices (i.e. a column times a row) where each next contribution in the sum is less important.
In mathematical parlance, any given matrix $A_{h\times d}$ can be written 
as: 
\begin{equation}
    A  = \sum\limits_{k = 1}^r \sigma_kU_k V_k^T  \quad \mbox{or more succinctly} \quad A = U S V^T
    \label{eq:svd}
\end{equation}
%     \quad\quad \mbox{or more succinctly} \quad A = USV^T
where $U \in {\cal O }(h)$  and $V  \in {\cal O }(d)$  are matrices with orthonormal columns, with $U_k$ and $V_k$ denoting the $k^{th}$ column of $U$ and $V$, respectively; and $S$ is an $h\times d$ matrix for which the only strictly positive elements $\sigma_k$ (so-called singular values) are situated on the main diagonal~\cite{baker2005singular}.
The importance of this decomposition lies in the fact that truncating the expansion in the right hand side of~\eqref{eq:svd} after the $p^{th}$ term yields the best  approximation of the original matrix $A$ by a matrix of (lower) rank $p\le r$~\cite{golub1970singular}. 
%It can be shown that $A_p$, reconstruction of order $p \leq r$ of matrix $A$, is closest (in $L_2$ norm) to the original matrix $A$:
\begin{equation}
     A_p = \arg\min\limits_{rank(R) = p} \n A-R \n 
\end{equation}
where the norm $ \n \cdot \n $ can be either the Frobenius or spectral $L_2$ norm. 
\begin{figure}
\centering
\hspace{-.8cm}
  \includegraphics[width=7.8cm,height=3.5cm]{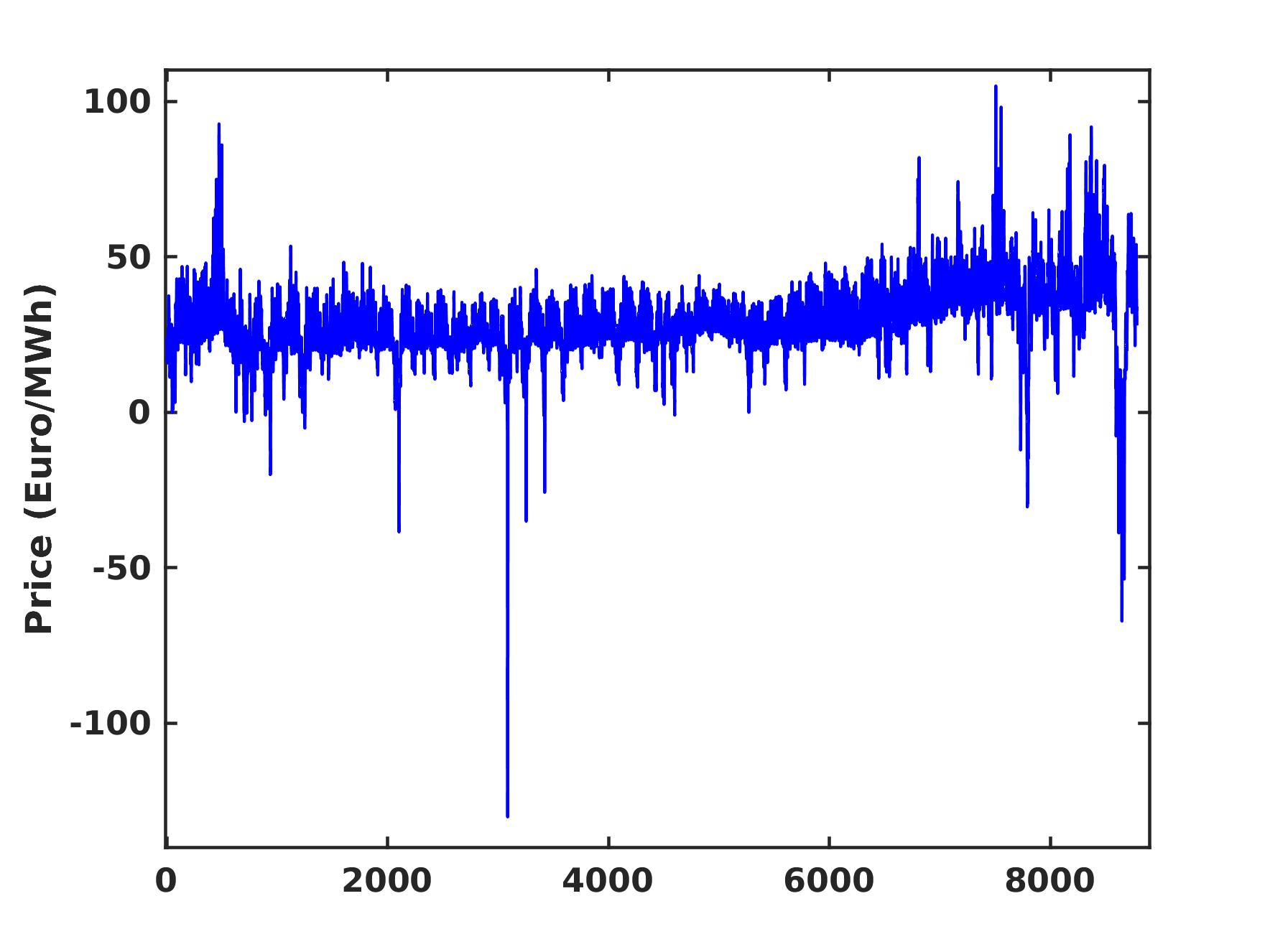}
  \caption{Raw price data for 2016}
  \label{fig:svd_price_raw}
\end{figure}
We illustrate the process for the price data of the year 2016. The raw  hourly values are shown  in Fig.~\ref{fig:svd_price_raw}. 
After recasting this time series into a $24 \times 366$ matrix (2016 was a leap year!), we can compute 
the first two terms in the SVD expansion. 
The results are illustrated in Fig.~\ref{fig:svd_rk2_approx}. 
The left panels depict the first two columns of the $U$ matrix. 
The top left panel resembles an appropriately weighted daily average--averaged over the year.
The price peaks in the morning and late afternoon are clearly visible.  
The second column of $U$ is depicted on the bottom left panel; it provides a correction which needs to be added to the average in order to get a better approximation. 
The panels on the right-hand side represent the first two columns of the $V$ matrix and specify an amplitude coefficient for every day. 
To obtain an approximation for the actual data on day $d$, one must multiply each profile on the left by the $d$-th amplitude coefficient on the right and of course the corresponding singular value; then add them all up. 
To get some idea of what a rank-2 approximation looks like, 
Fig.~\ref{fig:svd_rk2_approx_detail} shows a detail of  the approximation (in red) superimposed on the actual data (blue).
Fig.~\ref{fig:2016_price_residual} contains a more general case of the reconstruction of price data in 2016, along with the absolute value of the corresponding residuals. 
%and the first two terms in the SVD expansion 
%in Fig.~\ref{fig:svd_rk2_approx}
\subsection{Quantifying volatility}
\begin{figure}
    \centering
    \includegraphics[width=7.5cm,height=4cm]{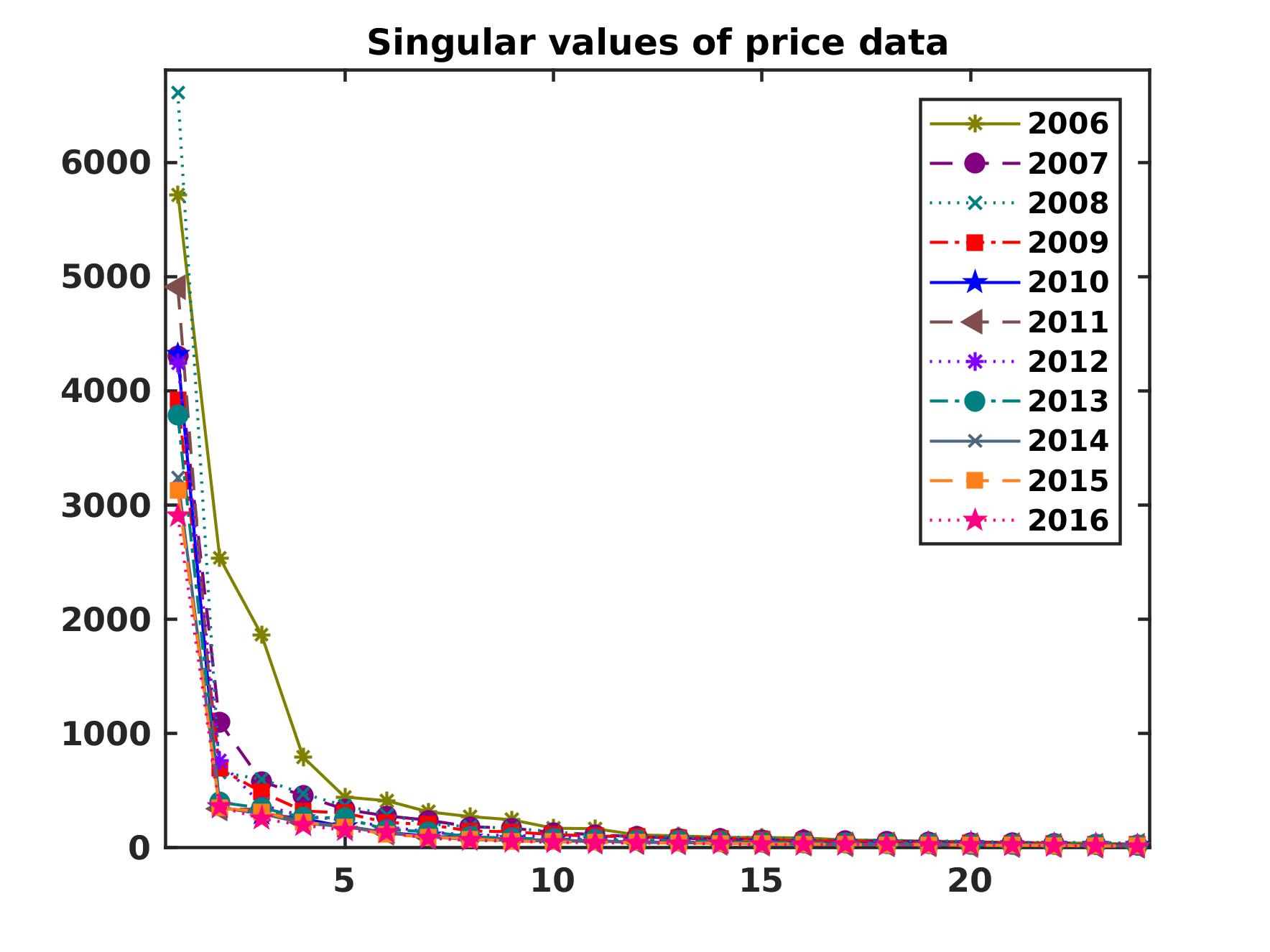}
    \caption{An overview of the evolution of the singular value of the price data in each year.}
    \label{fig:main_270916_GE_draft_fig88}
\end{figure}

\begin{figure}[]
\centering
  \includegraphics[width=8cm,height=5.25cm]{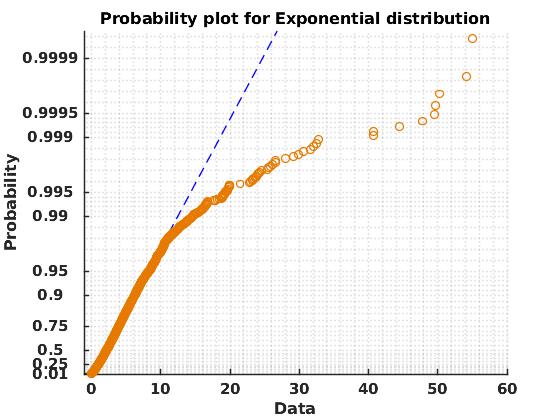}
  \includegraphics[width=8cm,height=5cm]{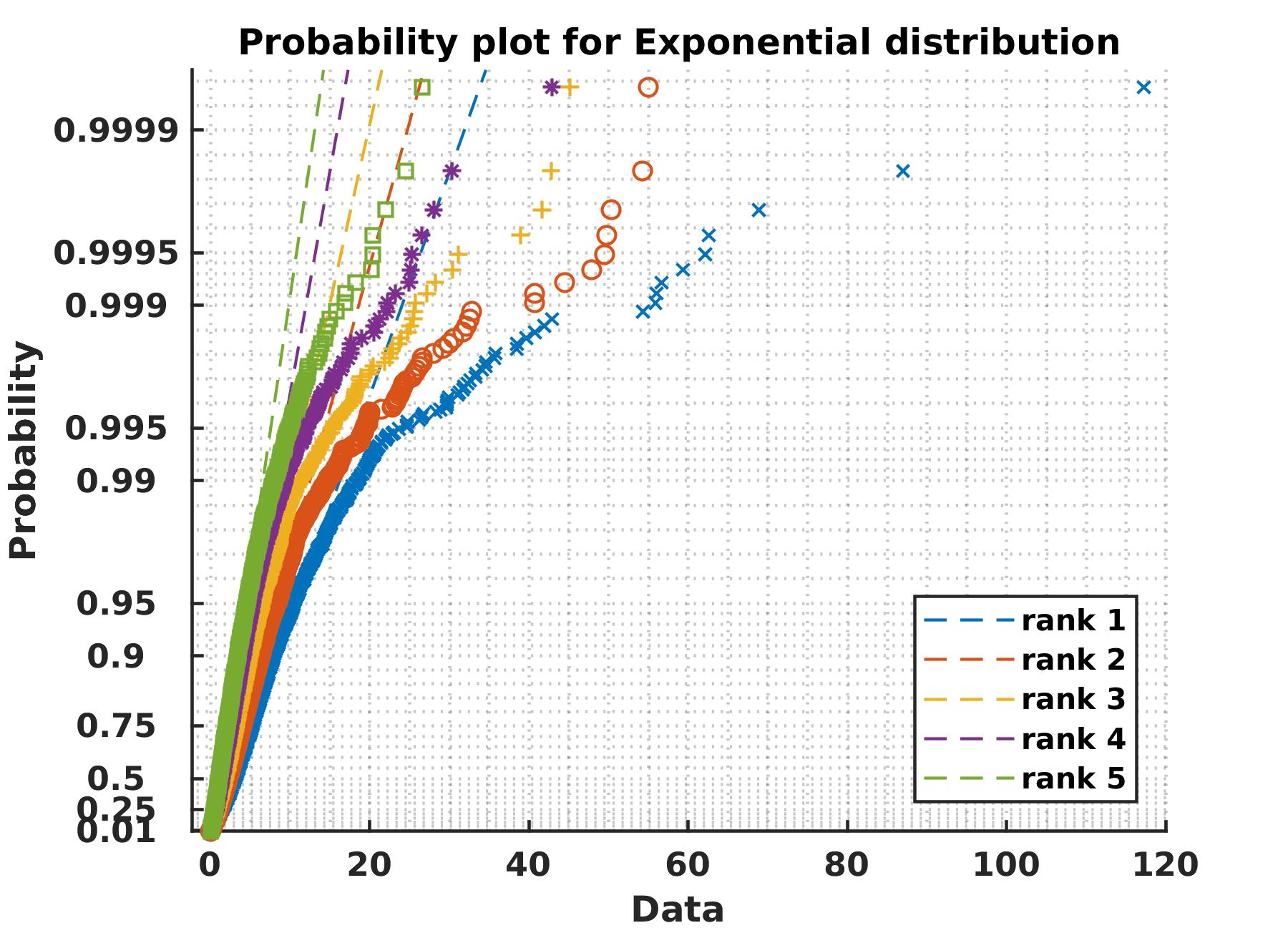}
  \caption{Top: The residuals of the rank-2 approximation of the day-ahead market prices for year 2016: almost 99\% of residuals adhere to exponential distribution. Bottom: Higher rank approximations yield similar results. }
  \label{fig:pr2016_exp_probplot}
\end{figure}

\begin{figure}
    \centering
    \includegraphics[width=8cm,height=5cm]{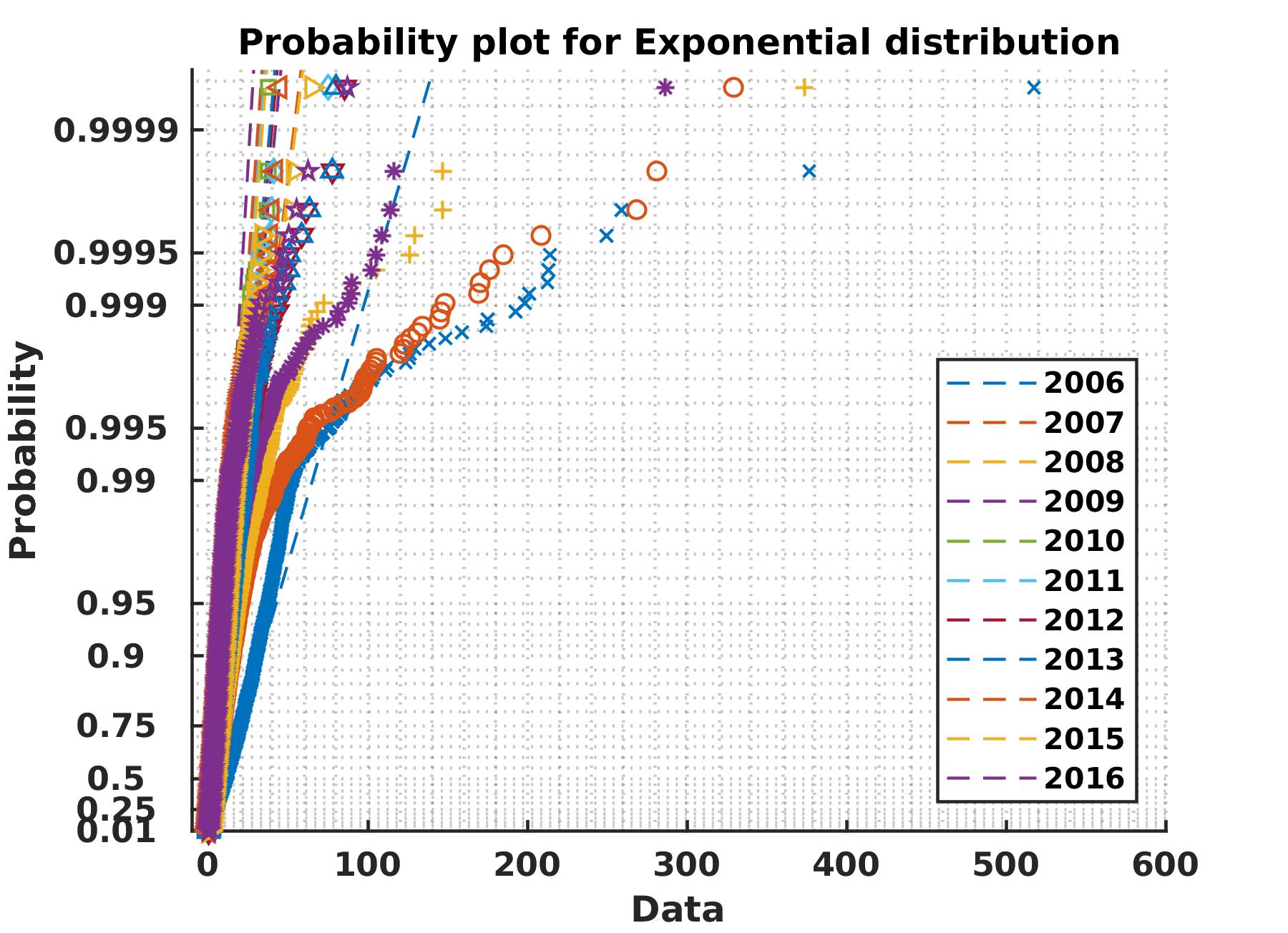}
    \caption{Exponential distribution of the residuals of the rank-2 approximations for different years.}
    \label{fig:main_270916_GE_draft_fig100}
\end{figure}

\begin{figure}
    \centering
    \includegraphics[width=8cm,height=5cm]{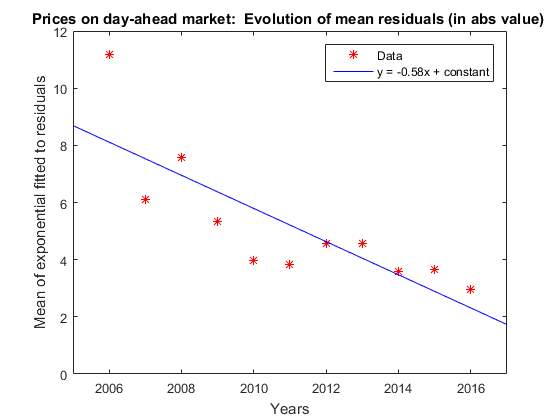}
    \caption{Evolution of the price volatility on the German day-ahead market in the period 2006 through 2016 .  
    The individual data points record the estimated exponential parameter (mean) based on all but the 1\% highest values of the residuals. The regression line has a slope equal to -0.58 which is highly significant (95\% confidence interval: -0.89\,:\,-0:26). See main text for more details.}
    \label{fig:pr_muhat_evolution_linregression}
\end{figure}

\begin{figure}[]
    \centering
    \includegraphics[width=8cm,height=5cm]{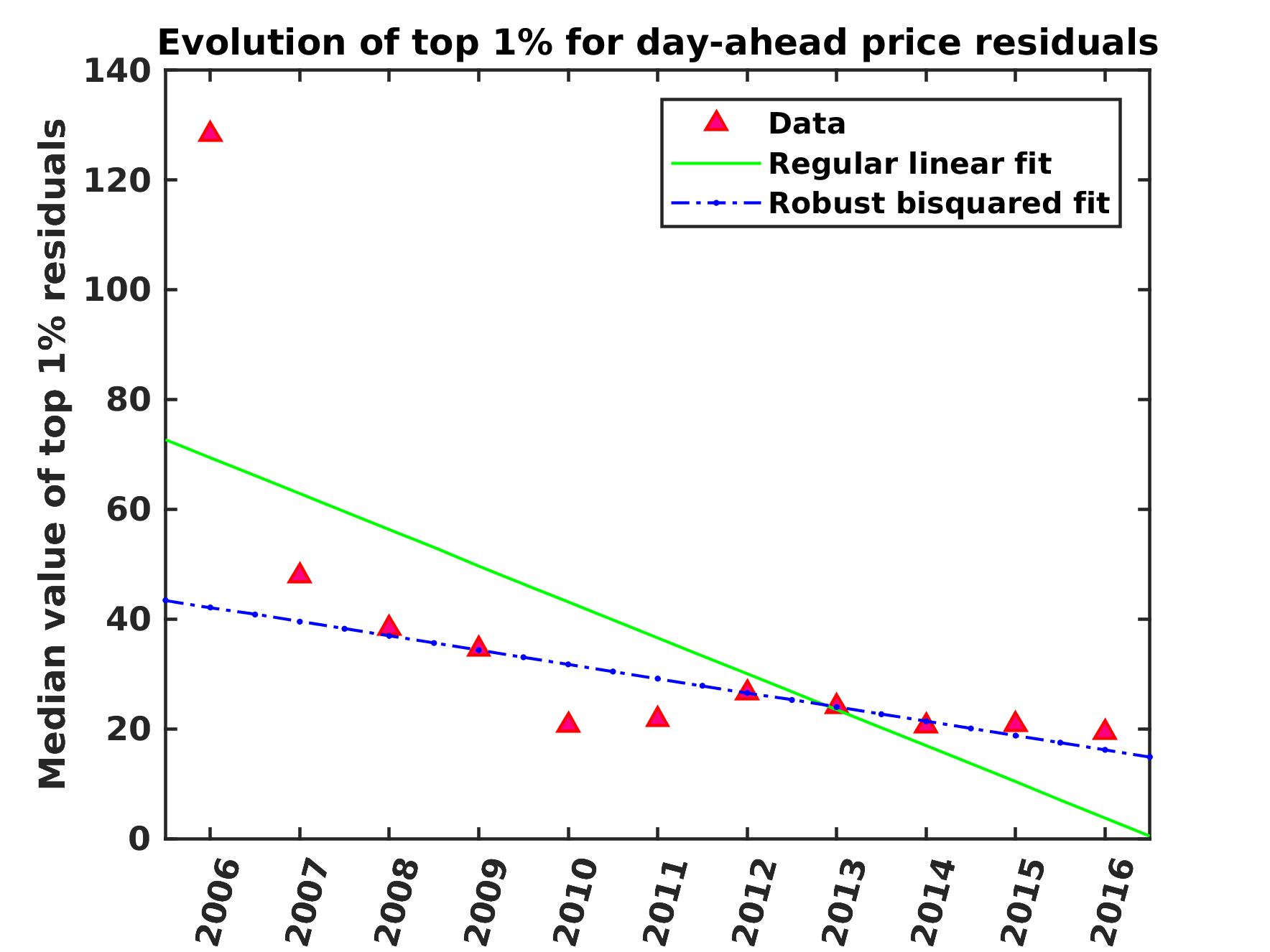}
    \caption{Evolution of top 1\% residuals over the years 2006-2016. Each data  point represents the median value of the top one percent residuals (in absolute value).  As such, these values characterize the extreme deviations in day ahead prices. The plot clearly shows that these extreme values decreased significantly before 2010, and then stayed approximately constant.}
    \label{fig:pr_topone_evolution}
\end{figure}

\begin{figure}
    \centering
    \includegraphics[width=8cm,height=5cm]{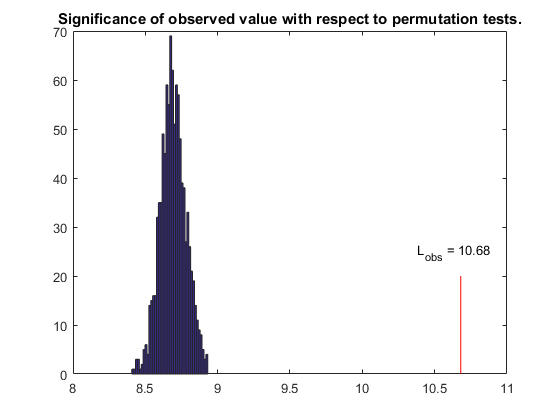}
    \caption{Price data for 2016:  Results of the permutation test. 
    The histogram of the $L$-values for 1000 random permutations of the actual data. 
    Obviously, the actually observed value (indicated in red)  is  significantly larger than values we would expect for data sets without structure
    (with a p-value $< 10^{-3}$).  This confirms our observation that volatility shows a seasonal pattern.}
    \label{fig:permutation_test}
\end{figure}

In the current section, we focus on the residuals of the price after compensating for daily patterns using a rank-2 approximation; it is done to quantify the volatility of the data over the years.
The choice to focus on rank-2 approximation is relatively arbitrary and unimportant. Fig.~\ref{fig:main_270916_GE_draft_fig88} illustrates that the singular values of the price data throughout different years follow the same pattern. 
%\textcolor{blue}{Higher rank approximation yield similar results.}
Furthermore, as can be seen in the top panel of Fig.~\ref{fig:pr2016_exp_probplot}, the absolute value of the residuals adhere remarkably well to an exponential distribution (with mean $2.97$). 
It is almost only the top 1\% that is substantially higher in value than expected. 
Moreover, the lower panel of Fig.~\ref{fig:pr2016_exp_probplot} highlights the fact that the higher rank approximations yield similar results.
Fig.~\ref{fig:main_270916_GE_draft_fig100} confirms that this pattern is also seen throughout the period (2006-2016) -- we focus on in this paper.
Based on this observation, we propose the following approach to quantify the evolution in (annual) volatility in the years 2006 through 2016: 
\begin{enumerate}
\item The influence of the daily and seasonal variations is removed by fitting a rank-2 approximation, and extracting the residual of the actual data 
with respect to this approximation.  As volatility is influenced by both positive and negative fluctuations, we focus on the absolute values of these residuals. 
\item For every year, we fit the lowest 99\% of the residual values with an exponential and compute the corresponding parameter (i.e. mean of the exponential).  This value corresponds to the size of the residuals. 
\item Typically, the top 1\% of the observed distribution is much larger than expected (based on the bulk of the distribution). 
We characterize these values by computing the median value this top 1\% segment separately.
\end{enumerate}
The results are shown in Figs.~\ref{fig:pr_muhat_evolution_linregression} and~\ref{fig:pr_topone_evolution}. 
The former figure shows a robust estimate for the mean of exponential distribution for each year.  
The estimate is based on the lowest 99\%  of the residual (absolute) values and is therefore robust with respect to the 1\% extremely large values. The 99\% vs. 1\% is dictated by the exponential prob-plot in Figs.~\ref{fig:pr2016_exp_probplot} and~\ref{fig:main_270916_GE_draft_fig100} which show a clear divergence at the 99\% mark. 
To quantify this decreasing trend, we have also computed  the regression line, which yields a statistically significant downward slope equal to  -0.58 (with 95\% confidence interval: 
-0.89:-0.26).  The quality  of this regression can be further improved by fitting a \textit{power law}, but at this point we simply want to illustrate the significant downward trend. 
The evolution of the top 1\% is shown in Fig.~\ref{fig:pr_topone_evolution} where these data are represented by their median value.  The picture we get from this indicates that whereas extreme residuals were not uncommon prior to 2010, these values fell significantly and have been roughly constant during the last five years. 
The message from both figures combined  is that volatility has decreased significantly over the past decade.
\paragraph{Volatility tends to be higher in winter}
Fig.~\ref{fig:2016_price_residual} suggests that volatility tends to be lower in summer (middle part of graph) than winter (extremal parts of the graph).
To demonstrate that this is indeed the case,  we use  a measure based on the angular momentum. More precisely,  if the residual for hour slot $h$ is given by $R(h)$ and the distance between the hour slot $h$ and the central hour slot $h_m = n/2 = 4392 $ equals $|h-h_m|$  the observed angular momentum is defined a: 
\begin{equation}
    L_{obs} = \sum_{h=1}^n R(h) (h-h_m)^2 
\end{equation}
where $n$ is equal to the total number of hour slots.
If we re-scale the values of the hour slot in such a way that $h_m=0$ and $-1 \leq h\leq 1$ (divided by 1000, for ease of comparison), 
we obtain $L_{obs} =10.676$.  A high value for $L_{obs}$  refutes the assumption that the residuals are uniformly distributed throughout the year, and favours an interpretation in which residuals (and hence volatility) is higher in the winter season. 
To judge the significance of this result, we use a permutation test. The rationale is straightforward: if the residuals are uniformly distributed over the hour slots, then a random permutation of the values should not result in a significantly different value for $L_{obs}$. 
The results for 2016 are depicted in Fig.~\ref{fig:permutation_test}.
It can be shown that all the other years produced similar results. 
\section{Conclusions }\label{sec:conclusion}
\begin{comment}
reviewer: 
\textcolor{blue}{
After accounting for seasonal and daily variations, it is concluded that variability has been decreasing. This trend is attributed to "market coupling has dampened the fluctuations and intermittencies inherent in renewables." This conclusion seems entirely misplaced to me. In the day-ahead market which is studied here, the renewable energy sources have not had a chance to act, intermittently or not. Instead, any variability in this market attributed to renewable energy sources must be due to errors in day-ahead projections for these sources.
It is a forecasting problem. In that light, a much more compelling conclusion would be that forecasting tools for renewables have increased in accuracy in the past 11 years, something that is certainly true. 
}
If the accuracy of the renewable power generation forecasting tools have increased substantially over the recent years?
Do we see their effect on the price volatility of the  day-ahead  market over the past eleven years (i.e. 2006-2016)? How can we quantify the volatility? 
On the other hand, market coupling 
across north-western Europe should have a stabilizing effect on prices. 
To gauge the relative effect of these opposing trends it is of interest to quantify 
the actual evolution of the price volatility. 
\end{comment}
The growing share of renewables (especially wind and solar) in the German energy mix appears to result in a higher degree of volatility of the electricity prices; as RES are intermittent and less predictable.
Any errors in day-ahead projections for these sources can lead to higher variability in the market.
In this paper, we looked at the electricity prices on the German day-ahead market in the period 2006 through 2016. 
The occurrence of zero or even negative prices makes it difficult to simply use the standard volatility measures which are common in fintech and stock markets. 
Therefore, we focused on residuals obtained after removing the underlying diurnal or seasonal patterns. 
This was accomplished by applying singular value decomposition (SVD) to the time series to get a low rank approximation of the raw data. 
Our results show that price volatility has significantly decreased during the past eleven years.  The most likely cause for this is the improved 
accuracy of the forecasting tools  for renewables  in recent years.

% or

\ifCLASSOPTIONcaptionsoff
  \newpage
\fi

\end{document}